\documentclass[12pt,preprint]{aastex}

\usepackage{graphicx}
\usepackage{txfonts}

\shorttitle{Small-scale, Dynamic Bright Blobs in Solar Filaments and Active Regions}
\shortauthors{Lin et al.}

\begin{document}

\title{Small-scale, Dynamic Bright Blobs in Solar Filaments and Active Regions}

\author{Y. Lin, O. Engvold and L.H.M. Rouppe van der Voort }
\affil{ Institute of Theoretical Astrophysics, University of Oslo, P.O. Box 1029, Blindern, N-0315
  Oslo, Norway}

\begin{abstract}
High cadence high spatial resolution observations in H$\alpha$ with the Swedish 1-m Solar Telescope on La Palma have revealed the existence of small-scale highly dynamic bright blobs. A fast wavelength tuning spectro-polarimeter provides spectral information of these structures. The blobs slide along thin magnetic threads at speeds in the range from 45 km\,s$^{-1}$ to 111 km\,s$^{-1}$. The blobs have a slight elongated shape and their lengths increase by a factor of 3 from close to $\frac{1}{2}$ arcsec when they first appear till they disappear 1$-$2 min later. The brightest blobs show the highest speed. The widths of the H$\alpha$ line emission of the blobs correspond to non-thermal velocities in the plasma less than 10 km\,s$^{-1}$ which imply that they are not the result of shock driven heating. The dynamic character of the bright blobs is similar to what can be expected from an MHD fast mode pulse.

\end{abstract}


\section{INTRODUCTION}

The combination of high spatial resolution and temporal cadence in modern optical solar telescopes has revealed richness in dynamic fine-scale chromospheric and coronal plasma structures. Telescopes in space such as the Hinode/Solar Optical Telescope (SOT) and the Solar Dynamics Observatory/Atmospheric Imaging Assembly (AIA) offer the huge advantage of seeing-free, long time series of homogeneous quality (e.g., \citealt{2011Natur.472..197B}). Ground-based solar telescopes reduce successfully seeing-induced deformation by adaptive optics systems (Swedish 1-m Solar Telescope, \citealt{2003SPIE.4853..370S}; The New Solar Telescope, \citealt{2010AN....331..636C}; \citealt{2011LRSP....8....2R}) in combination with image restoration methods (e.g., \citealt{2005SoPh..228..191V}). The advances in observational performance have opened new avenues in exploration of solar fine-structure (\citealt{2011Sci...333..316S}; \citealt{2011arXiv1112.0656A}).

In high cadence (3 frames s$^{-1}$) H$\alpha$ time series of two active regions obtained on 4 October 2005 with the Swedish 1-m Solar Telescope (SST; \citealt{2003SPIE.4853..341S}) on La Palma, \citet{2006ApJ...648L..67V} noticed groups of rapidly moving bright blobs. Some bright blobs were seen traveling at velocities ranging from 40 km\,s$^{-1}$ to 115 km\,s$^{-1}$ along magnetic loops. Other blobs appeared to be locked onto sideways moving fan-shaped systems of bright threads. The threads outline the fine-scale structure of the solar magnetic fields and demonstrate beyond doubt the magnetic association of the blobs. The images were all recorded in the H$\alpha$ line center and it was evident that more spectral information was required in order to grasp the true physical nature of the high speed blob events. In a time series in the Ca\,II\,H line at the solar limb obtained with Hinode/SOT, \citet{2010MNRAS.404L..74Z} noticed an upward propagating intensity blob with apparent propagation speed 35 km\,s$^{-1}$.

After several seasons running observing programs with imaging H$\alpha$ at high spatial and temporal cadence with SST, it has become clear that these high speed blob events are a rather common phenomenon in active regions. We present here the results from the analysis of the 4 data sets of active areas, where the fast moving blobs are readily seen. In the last data set, we utilized the CRISP Imaging Spectro-polarimeter (CRISP, \citealt{2008ApJ...689L..69S}) to obtain, for the first time, detailed spectroscopic observations of these elusive high speed solar features.

\begin{figure}[!t]     
\centering
\includegraphics[width=0.9\columnwidth]{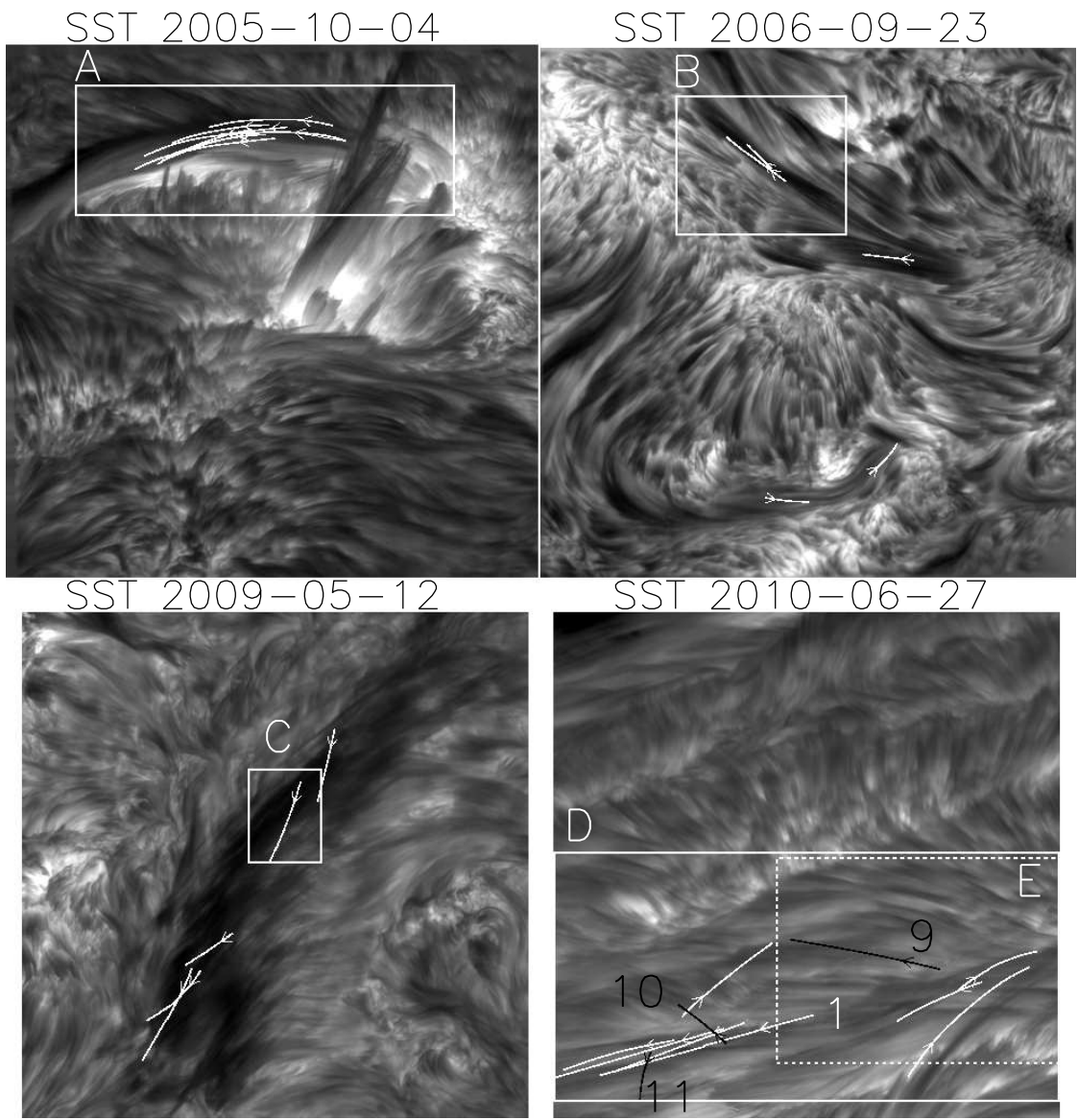}
\caption{Four H$\alpha$ line center filtergrams from four SST observing seasons. The size of the FOV in all panels is approximately 60$\arcsec\times$60$\arcsec$. 28 bright blobs were selected from these four data sets. The white lines indicate the trajectories of the fast moving bright blobs along the magnetic flux tubes. The dark ones mark the trajectories of the sideway moving threads on which some bright blobs are riding. The arrows show the directions of the bright blobs' motions. Movies of areas A$-$D (showing some of the high speed blobs along the magnetic flux tubes) and area E (showing some sideway moving blobs) are available in the online journal.}
\label{fig:SST_FOV_28HSB_pos}
\end{figure}

\begin{table}[htbp]
  \begin{center}
 \leavevmode
\small{
 \begin{tabular}{lcccc} \hline

 & { SST 2005-10-04} & { SST 2006-09-23} & { SST 2009-05-12} & { SST 2010-06-27} \\ \hline
 Position & S13W62 & S03W10 & N08E44 & S20E90\\
Time (UT)& 10:13:47-11:28:12 & 08:21:01-09:25:22 & 09:03:07-09:34:00 & 10:58:14-11:30:13 \\
Instrument & SOUP & SOUP & CRISP & CRISP \\
Cadence (s) & 1  & 1$^{*}$  & 1.3  & 9  \\
Data type & H$\alpha$ (0\AA) & H$\alpha$ (0\AA)  & H$\alpha$ (0\AA, $\pm$0.3\AA,  & H$\alpha$ (0{\AA} and 40 positions in \\
 & & & $\pm$0.45\AA) & the wing, 85 m{\AA} in step) \\ \hline
 \multicolumn{5}{l}{$^{*}$ The restored SST 2006 data was 1\,s cadence. However every 4th image was selected for the time series}\\
 \multicolumn{5}{l}{that is analyzed.}\\

 \end{tabular}}
  \end{center}
  \caption{A brief description of the four SST observations.}    
\label{tab:data_description}
\end{table}

A fifth CRISP/SST high-quality time series from 12 May 2009, 08:19:05UT $-$ 08:48:11UT, and cadence of 1.3 sec, was centered on a quiescent solar filament at position N14\,E33. No high speed bright blobs could be seen in this lower activity region which suggests tentatively that stronger magnetic fields may favor the generation of these dynamic small-scale features.

\section{OBSERVATIONS, INSTRUMENTATION AND DATA REDUCTION}
\label{sec_data}

The observations were obtained during 4 observing seasons at the Swedish 1-m Solar Telescope  on La Palma. Table~\ref{tab:data_description} summarizes some details and Figure~\ref{fig:SST_FOV_28HSB_pos} shows overview images of the different data sets. The FOV of the 2005 SST observations was centered on a decaying active region (AR 10812). At its border, a large active region filament was developed. Some dynamic surge-like structures can be seen in the center of the frame. This time series is one part of the data sets that were studied earlier by \citet{2006ApJ...648L..67V}. The FOV of the 2006 SST observations covered a small pore in AR 10910, that harbored dynamic fibrils (studied by \citealt{2008ApJ...673.1201L}) and active region filaments. The target of the 2009 SST observations was a section of an intermediate filament associated with an active region. The 2010 SST observing area was near a sunspot (AR11084) and close to SE limb. All targets from these four SST observational seasons are related to active regions. 

The data sets from the two earliest seasons were obtained with the Solar Optical Universal Polarimeter (SOUP, \citealt{1981siwn.conf..326T}) tunable filter of the Lockheed Martin Solar and Astrophysics Laboratory. This filter was then operated in single wavelength mode in the H$\alpha$ line center in order to achieve high temporal resolution. The two more recent data sets were obtained with the CRISP Imaging Spectro-polarimeter. CRISP is capable of high speed wavelength tuning ($<$50\,ms) and a detailed spectral line profile can be obtained in the course of a few seconds. For more details on the optical setup and observing strategies for the SOUP observations, we refer to \citealt{2007ApJ...655..624D}, and for CRISP observations we refer to \citealt{2009ApJ...705..272R}.

All observations took advantage of real-time tip-tilt corrections and adaptive-optics wave front corrections which are described by \citet{2003SPIE.4853..370S}. Post-processing of the raw images was done with the Multi-Object Multi-Frame Blind Deconvolution (MOMFBD) image restoration technique \citep{2005SoPh..228..191V}. Crucial for the achievement of high quality images at high temporal resolution is the use of high-speed cameras, we used CCD cameras at $\approx$ 37 frames s$^{-1}$. For the SOUP data sets, we used all 37 exposures in a single
MOMFBD restoration which resulted in a cadence of 1 s in the restored time series. For the CRISP data sets, we used 8 exposures per wavelength position so that 3 line positions were observed in less than 1\,s. It is important to point out the use of the wideband channel in the MOMFBD restoration process. A small fraction of the light is deflected from the main beam before the tunable filter instrument but after the prefilter. This beam holds a camera (a phase-diversity pair of cameras for the SOUP observations) that is synchronized to the narrowband cameras by means of an optical chopper. Precise alignment between the wideband and narrowband cameras is achieved by a separate alignment procedure involving a reference pinhole array target. For the MOMFBD restoration of the sequentially recorded CRISP images, the wideband channel serves as a so-called anchor channel that ensures accurate alignment between the different line positions. Despite the varying seeing conditions between sequentially recorded CRISP line positions, the spectral integrity for every spatial position in the FOV is guaranteed by the MOMFBD restoration process. This is particularly important for the study of low-contrast features moving at high speed. Furthermore, the restored wideband images serve as photospheric reference and are used to determine offsets and image deformations in the alignment and destretching process of the time series.

\section{PRESENTATION OF RESULTS}
\label{sec_results}
\subsection{Introduction}

An initial inspection of the four high-cadence high-spatial resolution time series showed that high speed bright blobs are commonly seen in H$\alpha$ images over active regions. Most often they appear in schools of single blobs sliding side by side at rather high speed along rather smooth trajectories as if they are generated by a common disturbance. Others appear to be tied to rapidly sideway moving thread-like structures (see the animations associated with Figure~\ref{fig:SST_FOV_28HSB_pos}, also available at http://folk.uio.no/yongl/Fast\_moving\_blobs\_movies.html).

In the following we select bright blobs that are associated with rather stable magnetic structures since these may better provide information about the plasma and dynamic of the blob per se. A total of 28 such blobs are selected from the four SST observing seasons. Their trajectories are indicated as the white lines in Figure~\ref{fig:SST_FOV_28HSB_pos}.

\begin{figure}[!t]     
\centering
\includegraphics[width=0.65\columnwidth]{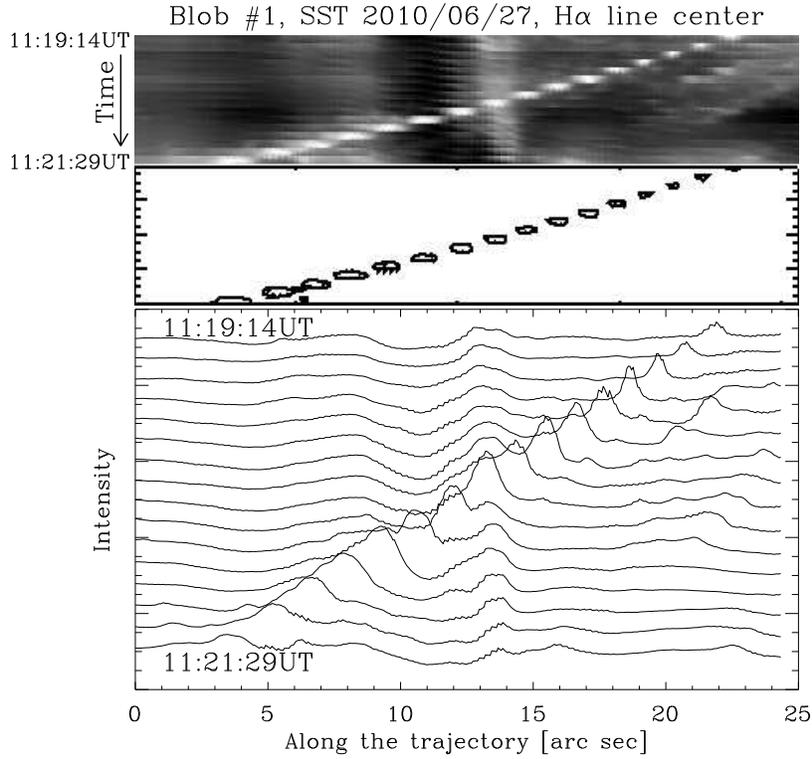}
\caption{Upper panel: The H$\alpha$ time slice diagram of blob \#1 from SST observations of 27 june 2010. The trajectory of this blob is indicated in the lower right panel of Figure~\ref{fig:SST_FOV_28HSB_pos}. Middle panel: Contour of this bright blob on the H$\alpha$ time slice diagram. One sees an increase in the length of the bright blob towards the end of the sequence. Lower panel: The excess intensity profiles of the blob along its trajectory in time series.}
\label{fig:SST2010June27_HSB1_int_curves}
\end{figure}

Time-slice diagrams are here used to display motion of small-scale solar structures in time series of images. Such diagrams are built from thin slices of the images oriented along the direction of motion of the feature at stake. By stacking side by side corresponding slices of the time series a moving feature will appear as an inclined streak in the resulting diagram. The steepness of the streak is inversely proportional to the speed of the feature in the plane of the sky (see upper part of Figure~\ref{fig:SST2010June27_HSB1_int_curves}).

\subsection{General characteristics of the features}

From the movies we get the impression that bright blobs are small single features. They are generally elongated with the long axis along the direction of motion. Their widths, typically $\sim$0.25$\arcsec$, are close to the resolution limit of the observations, while their lengths are from 2 to 8 times larger. The brighter and more accurately observed cases are seen to increase in length from about 0.6$\arcsec$ at their first appearance to close to 3 times larger by the time they disappear. One such case is illustrated in the time-slice diagram shown in Figure~\ref{fig:SST2010June27_HSB1_int_curves}. The observed intensity enhancement is restricted solely to the small bright bullet-like structure.

\begin{figure}[!t]     
\centering
\includegraphics[width=1.0\columnwidth]{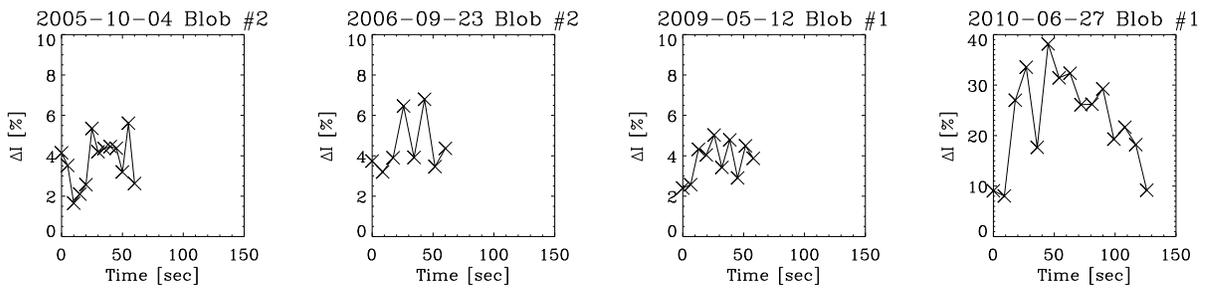}
\caption{The intensity enhancement of the 28 bright blobs were measured. As examples, four of them are shown here.}
\label{fig:SST_28HSB_I_measurement}
\end{figure}

The intensity of a bright blob is defined as its excess emission after subtraction of the local brightness along the trajectory and subsequently normalized to the mean intensity level of a nearby chromospheric area. The local brightness along the trajectory is taken as the mean of the intensities in either side of the trajectory close to the thin bright feature itself (see the sub-image in  Figure~\ref{fig:SST2010June27_HSB1_gaussianfit}). The blob intensities are thus found to vary notably within the sample of cases presented here. The weakest events are more easily detected against a darker background like in the case of the dark filament of 4 October 2005 (Figure~\ref{fig:SST_FOV_28HSB_pos}). 

\begin{figure}[!t]     
\centering
\includegraphics[width=0.65\columnwidth]{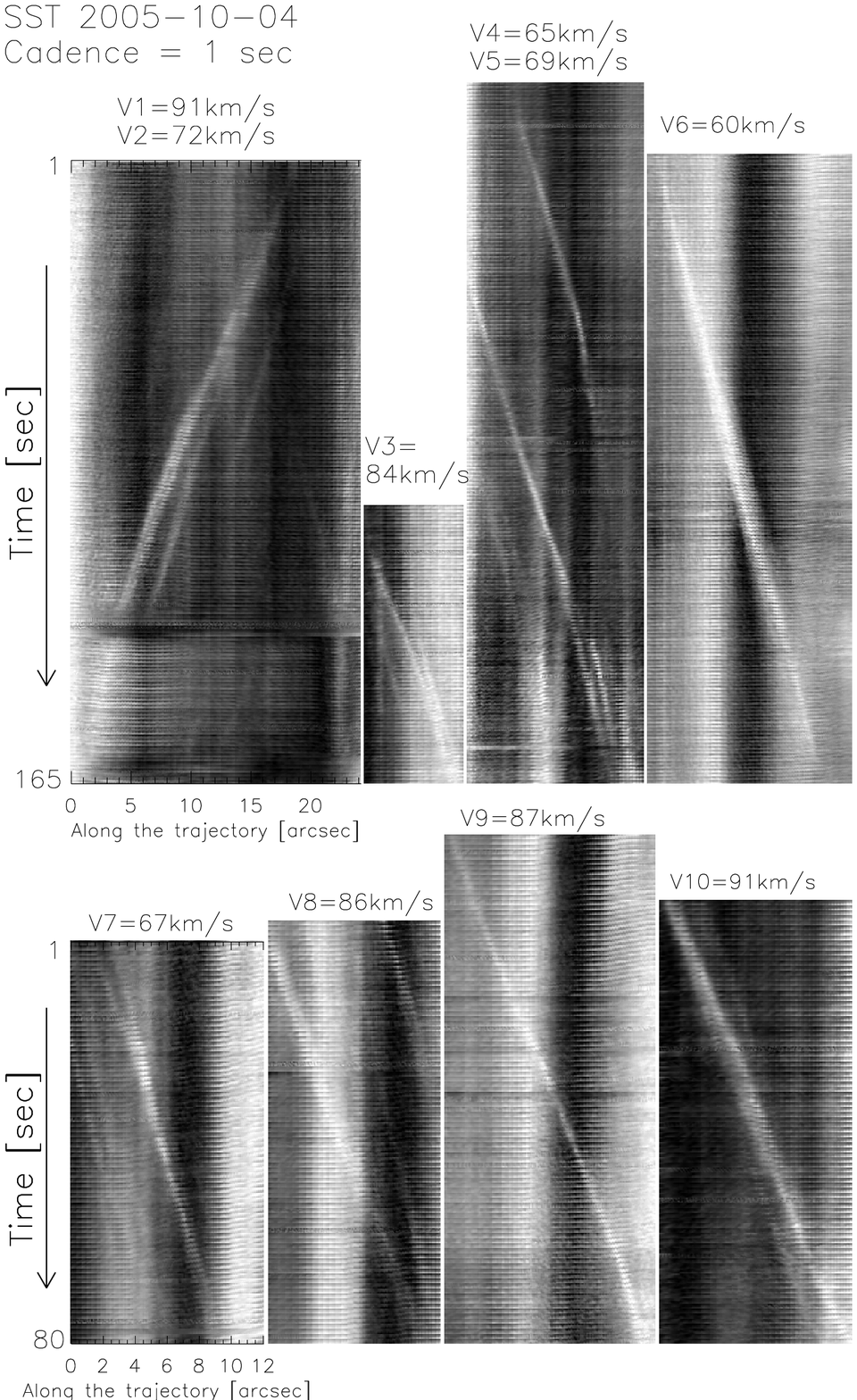}
\caption{H$\alpha$ time-slice diagrams of 10 fast moving bright blobs from SST 2005-10-04 observations.}
\label{fig:SST2005Oct04_10HSB_stackplot}
\end{figure}

\begin{figure}[!t]     
\centering
\includegraphics[width=0.65\columnwidth]{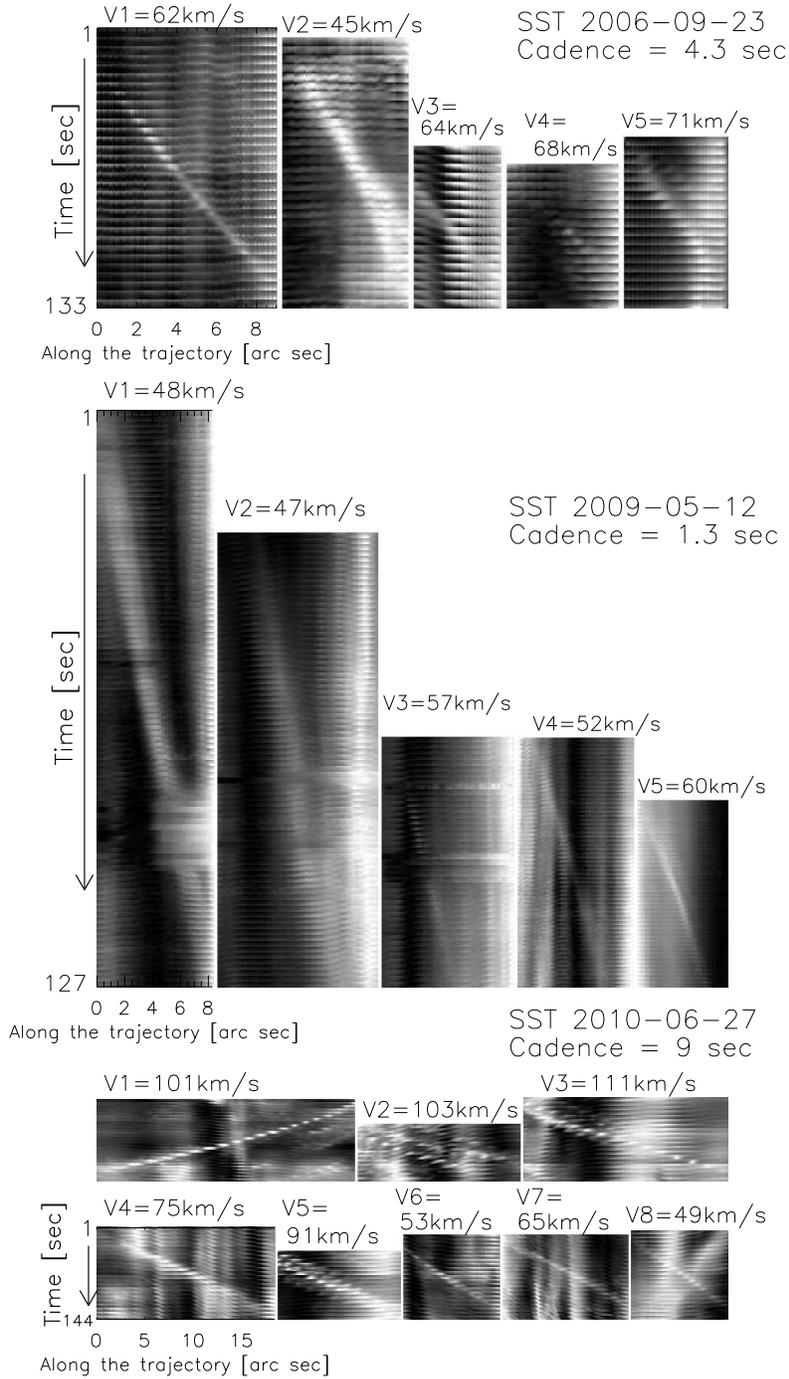}
\caption{H$\alpha$ time-slice diagrams of 18 fast moving bright blobs from SST 2006, 2009 and 2010 data sets.}
\label{fig:SST2006_2010_stackplot}
\end{figure}

The intensity variations of representative examples from the 28 bright blobs measured here are displayed in Figure~\ref{fig:SST_28HSB_I_measurement}. The fluctuation in intensity is notable, in particular for the weaker cases, which very likely reflect a combination of seeing variation and uncertainties in the derived local background intensities. In the following discussion one adopts the mean intensity values from the assumed relatively stationary time period.

Time-slice diagrams of 28 bright blobs are displayed in Figures~\ref{fig:SST2005Oct04_10HSB_stackplot} $-$~\ref{fig:SST2006_2010_stackplot}. Among these 28 cases, one finds that their speeds in the plane of sky range from 45 km\,s$^{-1}$ and 110 km\,s$^{-1}$. The speed of an individual feature exceeds the sound speed by a large factor and remains largely constant throughout their observed lifetime. No corrections for projection effects have been applied to these data implying that the derived speeds may be somewhat underestimated.

In the more agitated scenarios the bright blobs are seen to be locked onto sideways moving fan-shaped bundles of weakly emitting magnetic threads, whereas for blobs observed under more stable condition the threads are not observed directly but they are instead inferred from the generally straight smooth trajectories relative to the chromosphere below. The H$\alpha$ emission of the threads in the first group suggests that they are filled with cool plasma along their full length. The lack of visible thread structures for the latter group implies that the presumed cool plasma contained in these is optically thin and transparent in H$\alpha$.

The time-slice diagrams yield overall a constant speed of individual blobs. Some blobs show a slight change in speed in the course of their lifespan. The speeds of a very few blobs vary by up to 10 - 15\% (see Figures~\ref{fig:SST2005Oct04_10HSB_stackplot} $-$~\ref{fig:SST2006_2010_stackplot} and Table~\ref{tab:28HSBs}). Further studies may tell whether or not modest changes in speed in some bright blobs are an effect of thread geometry.

The recorded lifetime of a bright blob is the time from its first very weak appearance until it can no longer be seen. Their true lifetimes are very likely longer, but it cannot be determined from the present observations how much longer they last.

\subsection{Results from spectral data}
\label{sec:results_spectral_data}
\begin{figure}[!t]     
\centering
\includegraphics[width=0.7\columnwidth]{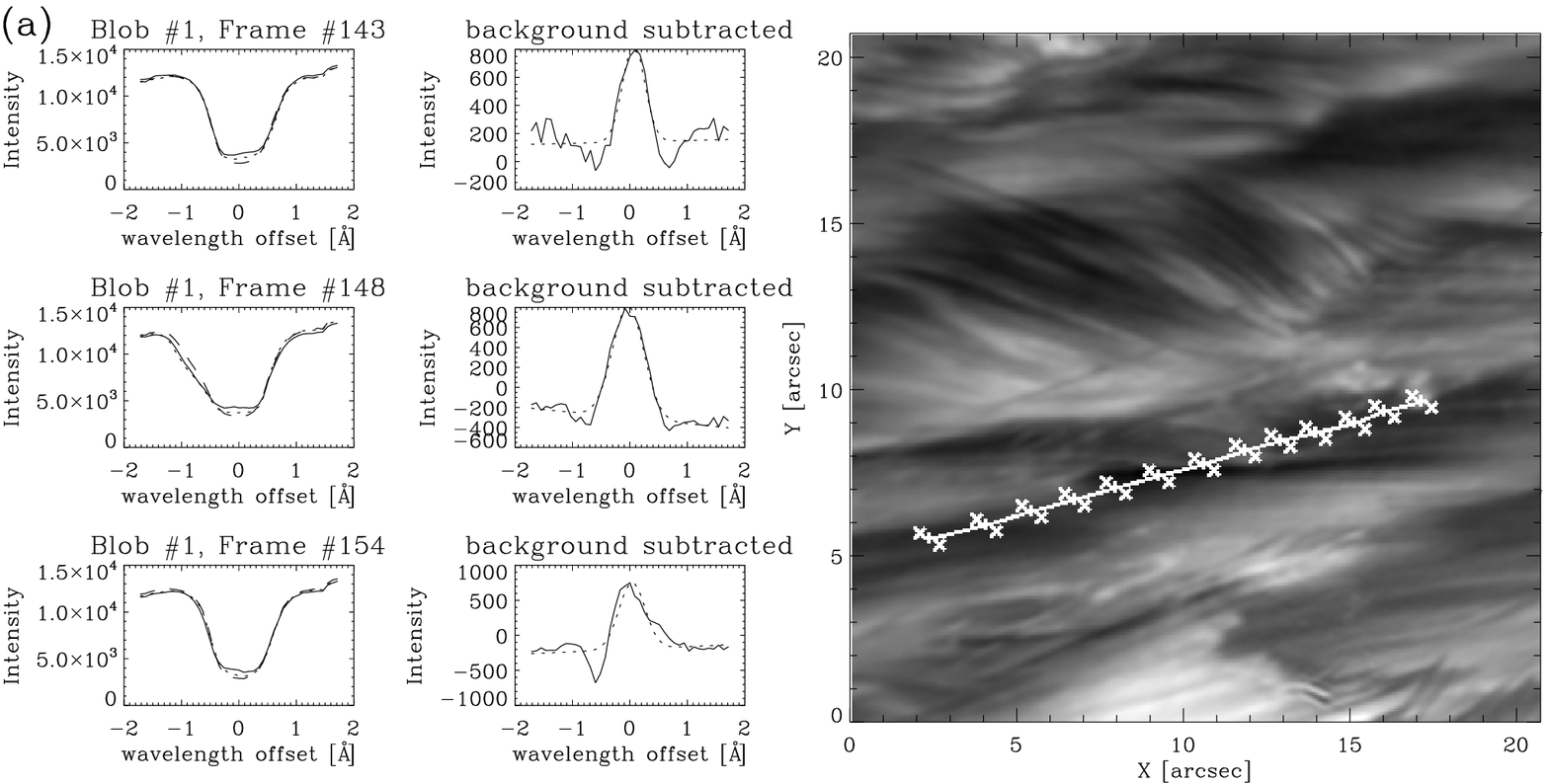}
\includegraphics[width=0.7\columnwidth]{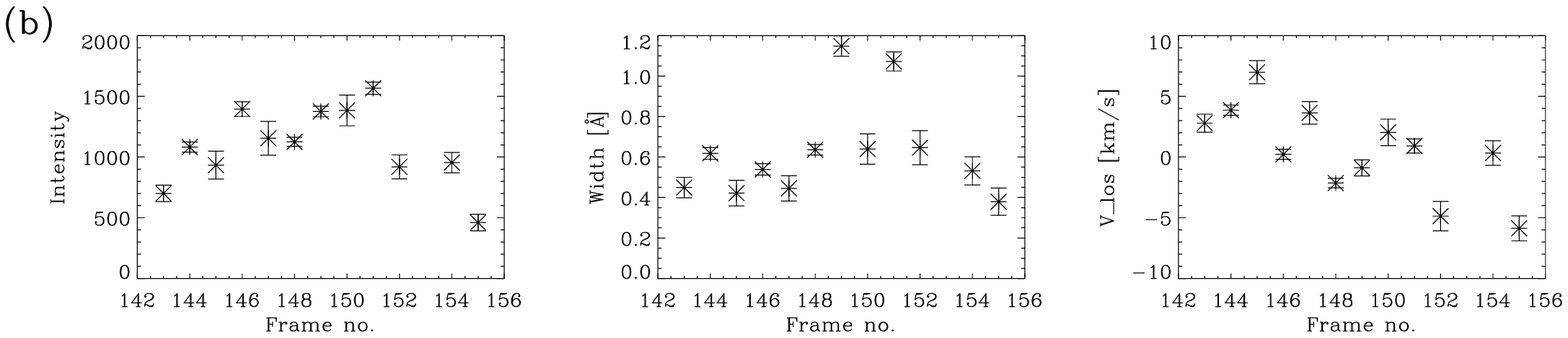}
\caption{A sub-image extracted from the lower left area of the SST 2010-06-27 filtergram (see Figure~\ref{fig:SST_FOV_28HSB_pos}) shows the trajectory of bright blob \#1 (solid line) from SST 2010 data set. Several pairs of points (marked as white crosses) on either side and close to the trajectory are used for local background measurements. The first column in panel (a) show the spectral profiles (solid lines) of blob \#1, in the beginning, middle and the end of its observable lifetime, superposed on the two local background profiles (dashed and dotted lines). The second column in panel (a) shows the emission line profiles (solid lines) after subtraction of the background. They are fitted by Gaussian function (dotted lines). Panel (b) shows the temporal variation of the derived intensity, width and V$_{los}$ from the fitted Gaussian curves.}
\label{fig:SST2010June27_HSB1_gaussianfit}
\end{figure}

The high speeds of bright blobs, in the plane of the sky, were derived earlier from single wavelength observations. To get a full dynamic picture of the blobs as well as of the internal kinetics of the blob plasma, one needs spectral information. Such information could be obtained from the time series of 27 June 2010. The first column in Figure~\ref{fig:SST2010June27_HSB1_gaussianfit}a shows spectral profiles of blob \#1 superposed on the profiles of the two spatial positions located close to the blob on either side of the trajectory. The mean value of these two line profiles is taken to represent the background profile in the trajectory onto which the blob emission is added. The second column of the same figure show the blob emission line profiles after subtraction of the adopted background profile. The resulting emission profiles are typically covered by about 9 wavelength steps which correspond to close to 2 seconds scan time. In the cases of the brighter ones among the blobs the emission profiles have a Gaussian
shape. It is noted that some emission profiles derived in this way deviates from a Gaussian profile. The method of subtracting an assumed underlying local line profile is subject to errors caused by spatial intensity variations in the background chromosphere.

The widths of the spectral line profiles thus obtained (e.g., Figure~\ref{fig:SST2010June27_HSB1_gaussianfit}b) yields line broadening velocities between 12 km\,s$^{-1}$ and 15 km\,s$^{-1}$. These line widths represent the combined thermal and non-thermal broadening for H$\alpha$ which are typical for solar prominences (\citealt{1995ASSL..199.....T}; \citealt{1998ASPC..150..175V}; \citealt{2010SSRv..151..243L}). 

\begin{figure}[!t]     
\centering
\includegraphics[width=0.8\columnwidth]{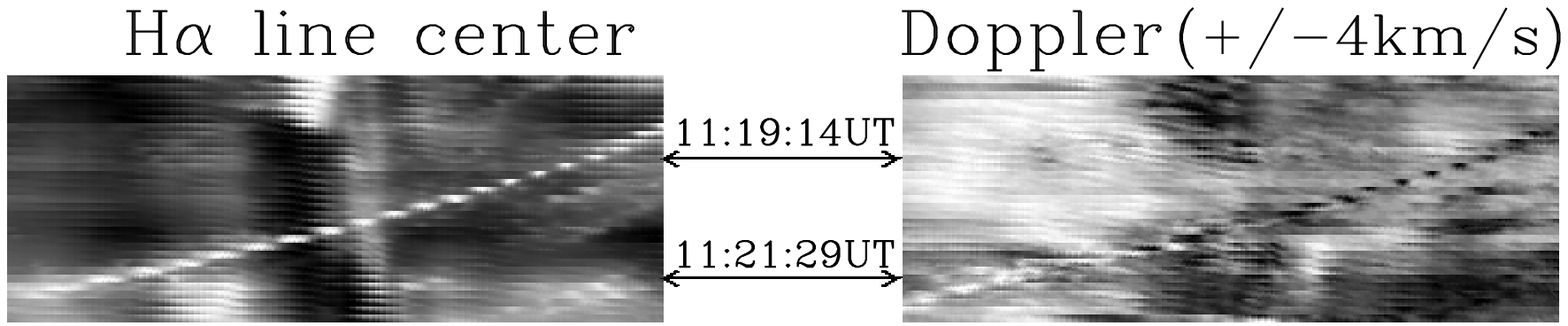}
\caption{The time-slice diagrams of blob \#1 (from SST 2010 June 27 observations) in H$\alpha$ line center and the corresponding H$\alpha$ Doppler. One notices that the Doppler signals of this bright blob changed its sign (from red shifted to blue shifted) during its observable period.}
\label{fig:SST2010June27_HSB1_timeslice_Ha_Doppler}
\end{figure}

The derived line emission of blobs all yield modest line shifts in the range of $\pm$5 km\,s$^{-1}$. The right panel of Figure~\ref{fig:SST2010June27_HSB1_gaussianfit}b and Figure~\ref{fig:SST2010June27_HSB1_timeslice_Ha_Doppler} show a sign change in the line-of-sight velocity within this range along a moving blob trajectory. This is most likely a result of a commonly occurring moderate swaying of the associated magnetic thread (cf., \citealt{2009ApJ...704..870L}).

\begin{table}[htbp]
  \begin{center}
 \leavevmode
 \begin{tabular}{cclll} 

{\bf Date} & {\bf Bright blob \#} & {\bf V} (km\,s$^{-1}$) & {\bf Lifetime} (second) & {\bf $\Delta$I (\%)} \\ \hline \hline

2005-10-04 & 1 &  90.9$\pm$1.3  & 98 & 5.8$\pm$1.8 \\ 
 & 2 &  	72.0$\pm$1.8   &     	62  & 3.7$\pm$1.2 \\
 & 	3 &  	84.3$\pm$0.9  &      	53  & 3.1$\pm$1.3 \\
 & 	4 &  	64.7$\pm$1.8  &      	70   & 2.1$\pm$2.9\\
 & 	5 &  	68.7$\pm$1.1   &     	116   & 2.7$\pm$1.4\\
 & 	6 &  	60.3$\pm$0.3   &     	151  & 5.6$\pm$3.6 \\
 & 	7 &  	67.1$\pm$1.0  &      	57  & 5.3$\pm$2.0 \\
 & 	8 &  	86.0$\pm$1.9 & 	70  &  3.4$\pm$1.7\\
 &  9 &  	87.2$\pm$1.3  &    	101  & 4.1$\pm$2.5\\
 & 	10  &  90.8$\pm$0.5 & 	88   & 4.2$\pm$1.8\\ \hline \hline

2006-09-23 & 1 & 61.6$\pm$0.7  & 86 & 0.6$\pm$2.1 \\ 
 & 2 & 44.6$\pm$4.0  & 82 & 4.5$\pm$1.4 \\ 
 & 3 & 64.4$\pm$1.4  & 55.9 & 2.6$\pm$1.1 \\ 
 & 4 & 67.9$\pm$19.7  & 34.4 & 4.1$\pm$2.0 \\ 
 & 5 & 70.5$\pm$10.2  & 38.7 & 6.5$\pm$2.9 \\ 
\hline \hline

2009-05-12 & 1 & 47.5$\pm$0.6  & 102  & 3.8$\pm$1.0 \\ 
           & 2 & 47.2$\pm$3.5  & 100  & 3.4$\pm$1.0 \\ 
           & 3 & 57.1$\pm$4.5  & 48  & 4.3$\pm$1.3\\ 
           & 4 & 52.0$\pm$2.1  & 70  & 2.2$\pm$1.3 \\ 
           & 5 & 60.0$\pm$1.3  & 46  & 4.3$\pm$1.7 \\ \hline \hline

2010-06-27 & 1 &  100.9$\pm$9.6  & 144  & 23.1$\pm$9.4 \\ 
 & 2 &  102.9$\pm$0.7 & 72  & 15.1$\pm$5.8\\ 
 & 3 &  111.0$\pm$11.6  & 108  & 14.0$\pm$6.2\\ 
 & 4 &  75.4$\pm$0.5  & 171   & 9.2$\pm$4.4 \\ 
 & 5 &  90.6$\pm$15.9 & 72   & 15.1$\pm$4.6\\ 
 & 6 &  53.3$\pm$0.4  & 108   & 5.4$\pm$2.3\\ 
 & 7 &  64.8$\pm$6.3  & 153  & 6.8$\pm$3.6 \\ 
 & 8 &  48.9$\pm$7.7 & 144 & 5.1$\pm$1.6\\ \hline 

 \end{tabular}
  \end{center}
  \caption{Several high speed blobs are selected from SST 2005-10-04 dataset (cadence of 1 sec), SST 2006-09-23 dataset  (cadence of 4.3 sec), SST 2009-05-12 dataset  (cadence of 1.3 sec) and SST 2010-06-27 dataset  (cadence of 9 sec). Their velocities in the plane of sky (V), observable lifetimes and intensity enhancement ($\Delta$I) were measured.}    
\label{tab:28HSBs}
\end{table}

\begin{figure}[!t]     
\centering
\includegraphics[width=0.7\columnwidth]{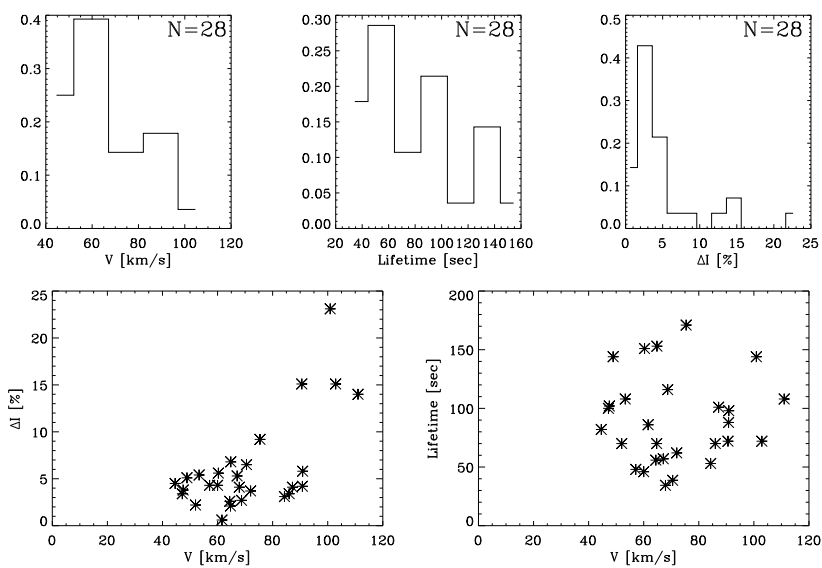}
\caption{Upper row:  Distribution of the measured velocities in the plane of the sky, lifetime and the intensity enhancement (\%) of the 28 fast moving bright blobs (see Table~\ref{tab:28HSBs}). Lower row: Relationships between the velocities in the plane of the sky and the two other parameters respectively.}
\label{fig:SST2005_2010_Vps_I_lifetime_summary}
\end{figure}

\subsection{Statistical results}
The derived characteristics of the 28 high speed bright blobs from the four time series are summarized in Table~\ref{tab:28HSBs}. With the exception of a few cases, the speeds could be determined quite accurately. The derived intensities relative to the mean background chromosphere show notable fluctuation around the mean values for reasons discussed above. 

Figure~\ref{fig:SST2005_2010_Vps_I_lifetime_summary} (lower row) shows that the fastest moving blobs also are the brighter ones. On the other hand, there is no apparent relation between their measured lifetimes and speeds.

\subsection{The behavior of blobs associated with sideway moving thread-like features}

The data sequence of 27 June 2010 contains several bright blobs that are clearly associated with agitated sideway moving thread-like structures (see the animation associated with Figure~\ref{fig:SST_FOV_28HSB_pos} area E).

\begin{figure}[!t]     
\centering
\includegraphics[width=1.0\columnwidth]{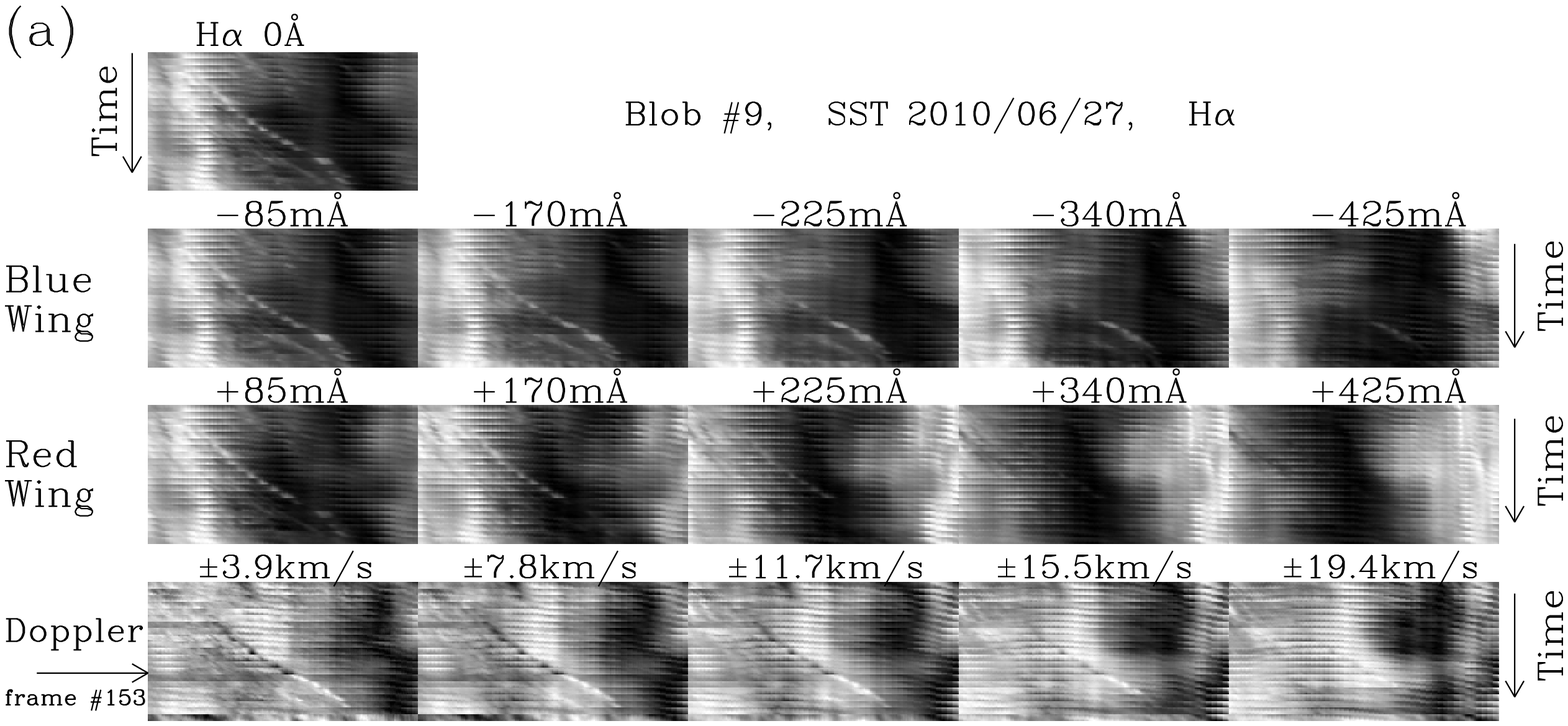}
\includegraphics[width=1.0\columnwidth]{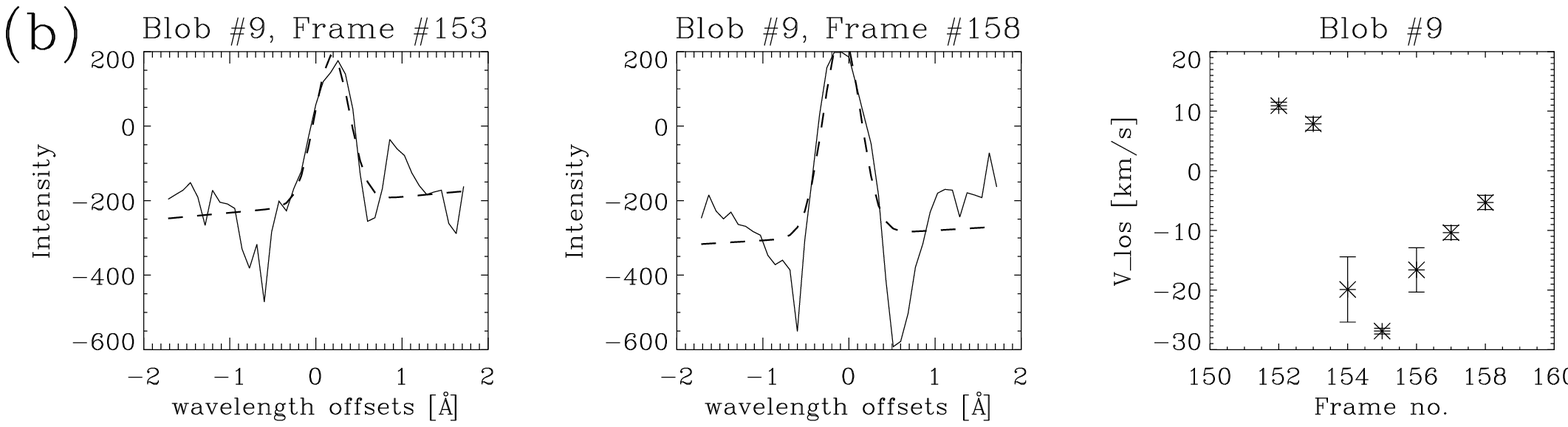}
\caption{Blob \#9 from SST 2010 observations moves along with a sideway moving thread, whose position is marked in the lower right panel of Figure~\ref{fig:SST_FOV_28HSB_pos}. Panel (a): The time slice diagrams of blob \#9 in H$\alpha$ line center, five blue wing, five red wing positions and the corresponding Dopplergrams, respectively. Panel (b): The excess spectral emissions of the blob (the solid lines) in frame \#153 and  \#158. The dashed lines show the applied Gaussian fits. The temporal variation of the Doppler velocities of this blob is shown in the last frame.}
\label{fig:SST2010June27_HSB9_sidewaymotion}
\end{figure}

Figure~\ref{fig:SST2010June27_HSB9_sidewaymotion}a shows the time-slice diagrams from one of these blobs (i.e., blob \#9). Doppler signals of the same feature are derived from subtracting time-slice diagrams from images of opposite wavelength positions in the line wing symmetric about the line center and subsequently normalized to the sum of these two wing images (see the left and middle panel of Figure~\ref{fig:SST2010June27_HSB9_sidewaymotion}b).

A noted change in the inclination of the Doppler signal diagram corresponds to a sideway speed of 57 km\,s$^{-1}$ in the first half of the sequence while it becomes 72 km\,s$^{-1}$ in the second half period. A comparison of the time-slice diagrams derived from 5 wavelength positions in the blue and the red wing of H$\alpha$, respectively, indicates a notable Doppler shift of the bright blob seen in frame \#153 of the time series. The derived emission line of the same blob is redshifted by about +8 km\,s$^{-1}$. The corresponding emission line from frame $\#$158 in the second half of the sequence yields a line of sight velocity of $-$5 km\,s$^{-1}$. The diagram in the rightmost panel of Figure~\ref{fig:SST2010June27_HSB9_sidewaymotion}b shows a notable variation in emission line shift from 7 frames in the time sequence ranging from $+$11 km\,s$^{-1}$ to $-$27 km\,s$^{-1}$. These changes in velocity suggest that this particular bright blob is connected with a rapidly moving bundle of thread-like structures and the motion change direction in the course of the sequence. The measured blob speed is most likely a combination of respectively the thread motion and the blob motion along the thread. 

The speeds of two other sideways moving blobs (denoted blobs \#10 and \#11 in Figure~\ref{fig:SST_FOV_28HSB_pos}) are 66\,km\,s$^{-1}$ and 63\,km\,s$^{-1}$, respectively. The line-of-sight velocities of these two blobs are within $-$5\,km\,s$^{-1}$ and $+$5\,km\,s$^{-1}$, which suggests that the motions of the two associated threads are mainly in the plane of the sky.

The sideway moving bright threads previously described by \citet{2006ApJ...648L..67V}, showed speeds in the range of 100$-$200\,km\,s$^{-1}$. The higher speeds could possibly represent the combined speeds of the bright blob and the thread or conceivably be associated with stronger magnetic fields.

\section{ALTERNATIVE EXPLANATIONS FOR THE DYNAMIC BRIGTH BLOBS}
A complete theory for high speed bright blobs must explain
\begin{itemize}
\item why they appear as single features
\item why they individually travel at nearly constant speed in the range from 45 km\,s$^{-1}$ to 111 km\,s$^{-1}$
\item why they often travel in schools of individual blobs
\item why the faster blobs are brighter
\item why their line emission yields non-thermal velocities less than 9 km\,s$^{-1}$
\item why they show a systematic slight increase in length with time
\item why they last for only 1$-$2 minutes  
\end{itemize}

Given that the typical length of the blobs are 400 $-$ 1200 km and their speeds are in the range given above, it implies that the cause of enhanced emission exist for only a very few seconds, up to a maximum of 20s, at a given location along bright blob trajectories. The $\sim$2 seconds scan time for the line emission of the blobs is thereby a fraction of the passage times of the blobs. Conceivably the changes in length of some bright blobs might also be associated with variation of the magnetic fields with height in the associated threads.

The fact that activated scenarios show some bright blob-associated threads in H$\alpha$ suggests that they are filled with cool plasma. Hence, explanations and models for high speed bright blobs should obviously involve chromospheric or prominence-like low-$\beta$ plasma. Magnetosonic waves and/or pulses are strongly influenced by the small-scale magnetic flux tubes which will act as wave guides. The following parameter values for chromospheric and prominence-like plasma are therefore relevant in a preliminary discussion on the physical nature of the blobs; Temperature T$_{e}$ = 7\,500K $-$ 10\,000K, number density N (cm$^{-3}$) from 2$\times$10$^{10}$ to 2$\times$10$^{11}$, plasma density $\rho$ (g cm$^{-3}$) from 2$\times$10$^{-14}$ to 2$\times$10$^{-13}$, and magnetic field strength B between 2 and 10 gauss (cf. \citealt{1995ASSL..199.....T}; \citealt{2004soas.book.....F}).

The sound speed c$_{s}$ for the adopted parameter values range between 11 and 13 km\,s$^{-1}$, whereas the Alfv\'en speed (V$_{A}$) varies from 20 km\,s$^{-1}$ to more than 300 km\,s$^{-1}$. MHD slow-mode $\alpha_{slow}$ and fast-mode $\alpha_{fast}$ wave speeds are derived from their well known relation to sound speed and Alfv\'en speed. The resulting value for the slow-mode $\alpha_{slow}$ varies from 5.5 km\,s$^{-1}$ to 6 km\,s$^{-1}$ whereas $\alpha_{fast}$ exceeds the corresponding Alfv\'en speed by a very small amount.

Consequently, the high speeds of bright blobs can not be accounted for as a sound pulse and neither as MHD slow-mode $\alpha_{slow}$. Similarly, an order of magnitude estimate of the propagation speed of a thermal perturbation under conditions of the plasma adopted above also appears much too small to account for high speed bright blobs \citep{2011private..communication}.

It appears most unlikely that the bright blobs represent mass flows since they move 3$-$7 times faster than the local velocity of sound. The rather moderate internal velocities of the blob plasma which are derived from the line emission of a number of blobs also seem to exclude that the fast moving blobs are due shock driven heating of the plasma. Given a number density of N$_{e}$ $\le$ 10$^{11}$cm$^{-3}$ in a partly ionized hydrogen gas one should expect cooling time of the order of minutes \citep{1978SoPh...56...87E} which clearly should give rise to a bright blob tail. That is not observed.

Radiation in the Balmer series of hydrogen is controlled by the recombination rate via the electron density of the plasma and becomes proportional to N${_e}^{2}$. The intensity enhancement of high speed bright blobs might therefore possibly result from a compression of the plasma. The measured enhanced emission of the blobs relative to the chromosphere background in the H$\alpha$ line center vary from around 10\% up to about 40\% in one exceptionally bright case. Such intensity enhancements would imply compressions of the gas, and hence compression of the magnetic field as well, ranging from about 5\% to maximum 18\%. A pure Alfv\'enic wave pulse will not lead to compression of the gas whereas a fast-mode wave pulse is compressive.

\citet{2010MNRAS.404L..74Z} suggest that the upward moving feature observed in the Ca\,II\,H line at the solar limb could represent a slow sausage soliton (see \citealt{1982MNRAS.198P...7R}; \citealt{2003A&A...404..701B}) propagating along a magnetic flux tube. It has been demonstrated that non-linear sausage waves may move faster than the sound speed (cf. \citealt{2003.book.....R}), but further studies are necessary to decide whether or not the observed characteristics the 28 bright blobs studied here can be accounted for as MHD solitons.

\section{CONCLUDING REMARKS}
\label{sec_conclusions}
The present study of the four H$\alpha$ time series of high cadence and high spatial resolution provides new information on small-scale highly dynamic bright blobs in both filaments and chromospheric fibrils related to active regions. During their transient appearance with lifetimes of $\sim$ 1-2 min the blobs are most often seen sliding along invisible smooth trajectories, while occasionally they are tied to rapidly moving bright threads. There is no evidence in the present data for the dynamic bright blobs being associated with the notably lower velocity mass flows commonly observed in solar filaments and in active regions. Their speeds along thread-like magnetic structures and their spectral characteristics seem to exclude an origin associated with acoustic or slow mode MHD wave pulses. Neither do the dynamic blobs have the character of pure thermal pulses. More detailed investigations will be needed to decide whether or not high speed bright blobs represent MHD fast-mode wave pulses or other mechanisms.

\begin{acknowledgements}
The authors like to thank R. Oliver, J.-L. Ballester, S.\,F. Martin and H. Pecseli for fruitful discussions in the course of this work. We are most grateful to the staff of the SST for their invaluable support with the observations. The Swedish 1-m Solar Telescope is operated on the island of La Palma by the Institute for Solar Physics of the Royal Swedish Academy of Sciences in the Spanish Observatorio del Roque de los Muchachos of the Instituto de Astrof{\'\i}sica de Canarias. We acknowledge Michiel van Noort, Ragnvald J. Irgens, Dan Henrik Sekse and Patrick Antolin for co-observing of SST 2005, 2006, 2009 and 2010 data sets, respectively. We also thank Gregal Vissers for helping with CRISPEX software during the data analyses and Robertus Erd\'elyi for useful discussions and comments.

\end{acknowledgements}

\clearpage


\begin{thebibliography}{}

\bibitem[{Antolin} \& {Rouppe van der Voort}(2011)]{2011arXiv1112.0656A}
{Antolin}, P. \& {Rouppe van der Voort}, L. (2011).
\newblock {\apj}, {\em ArXiv e-prints}.

\bibitem[{Ballai} et al.(2003)]{2003A&A...404..701B}
{Ballai}, I., {Thelen}, J.~C., \& {Roberts}, B. (2003).
\newblock {\aap}, {404}, 701.

\bibitem[{Berger} et al.(2011)]{2011Natur.472..197B}
{Berger}, T., {Testa}, P., {Hillier}, A., {Boerner}, P., {Low}, B.~C.,
  {Shibata}, K., {Schrijver}, C., {Tarbell}, T., \& {Title}, A. (2011).
\newblock {\nat}, {472}, 197

\bibitem[{Cao} et al.(2010)]{2010AN....331..636C}
{Cao}, W., {Gorceix}, N., {Coulter}, R., {Ahn}, K., {Rimmele}, T.~R., \&
  {Goode}, P.~R. (2010).
\newblock {Astronomische Nachrichten}, {331}, 636

\bibitem[{De Pontieu} et al.(2007)]{2007ApJ...655..624D}
{De Pontieu}, B., {Hansteen}, V.~H., {Rouppe van der Voort}, L., {van Noort},
  M., \& {Carlsson}, M. (2007).
\newblock {\apj}, {655}, 624

\bibitem[{Engvold}(1978)]{1978SoPh...56...87E}
{Engvold}, O. (1978).
\newblock {\solphys}, {56}, 87

\bibitem[{Foukal}(2004)]{2004soas.book.....F}
{Foukal}, P.~V. (2004).
\newblock Wiley-VCH Verlag GmbH \& Co.

\bibitem[{Labrosse} et al.(2010)]{2010SSRv..151..243L}
{Labrosse}, N., {Heinzel}, P., {Vial}, J.-C., {Kucera}, T., {Parenti}, S.,
  {Gun{\'a}r}, S., {Schmieder}, B., \& {Kilper}, G. (2010).
\newblock {\ssr}, {151}, 243

\bibitem[{Langangen} et al.(2008)]{2008ApJ...673.1201L}
{Langangen}, {\O}., {Rouppe van der Voort}, L., \& {Lin}, Y. (2008).
\newblock {\apj}, {673}, 1201

\bibitem[{Lin} et al.(2009)]{2009ApJ...704..870L}
{Lin}, Y., {Soler}, R., {Engvold}, O., {Ballester}, J.~L., {Langangen}, {\O}.,
  {Oliver}, R., \& {Rouppe van der Voort}, L.~H.~M. (2009).
\newblock {\apj}, {704}, 870

\bibitem[{Oliver}(2011)]{2011private..communication}
{Oliver}, R. (2011).
\newblock {private communication}.

\bibitem[{Ruderman}(2003)]{2003.book.....R}
{Ruderman}, M.~S. (2003).
\newblock in Erd\'elyi R. et al., eds, Turbulence, Waves, and Instabilities in the Solar Plasma, Kluwer, Dordrecht, 239

\bibitem[{Rimmele} \& {Marino}(2011)]{2011LRSP....8....2R}
{Rimmele}, T.~R. \& {Marino}, J. (2011).
\newblock {\em Living Reviews in Solar Physics}, {8}, 2

\bibitem[{Roberts} \& {Mangeney}(1982)]{1982MNRAS.198P...7R}
{Roberts}, B. \& {Mangeney}, A. (1982).
\newblock {MNRAS}, {198}, 7.

\bibitem[{Rouppe van der Voort} et al.(2009)]{2009ApJ...705..272R}
{Rouppe van der Voort}, L., {Leenaarts}, J., {de Pontieu}, B., {Carlsson}, M.,
  \& {Vissers}, G. (2009).
\newblock {\apj}, {705}, 272

\bibitem[{Scharmer} et al.(2003a)]{2003SPIE.4853..341S}
{Scharmer}, G.~B., {Bjelksjo}, K., {Korhonen}, T.~K., {Lindberg}, B., \&
  {Petterson}, B. (2003a). Proc. SPIE, 4853, 341

\bibitem[{Scharmer} et al.(2003b)]{2003SPIE.4853..370S}
{Scharmer}, G.~B., {Dettori}, P.~M., {Lofdahl}, M.~G., \& {Shand}, M. (2003b).
Proc. SPIE, 4853, 370

\bibitem[{Scharmer} et al.(2008)]{2008ApJ...689L..69S}
{Scharmer}, G.~B., {Narayan}, G., {Hillberg}, T., {de la Cruz
  Rodr{\'{\i}}guez}, J., {L{\"o}fdahl}, M.~G., {Kiselman}, D., {S{\"u}tterlin},
  P., {van Noort}, M., \& {Lagg}, A. (2008).
\newblock {\apjl}, {689}, L69

\bibitem[{Scharmer} et al.(2011)]{2011Sci...333..316S}
{Scharmer}, G.~B., {Henriques}, V.~M.~J., {Kiselman}, D., \& {de la Cruz
  Rodr{\'{\i}}guez}, J. (2011).
\newblock {Science}, {333}, 316

\bibitem[{Tandberg-Hanssen}(1995)]{1995ASSL..199.....T}
{Tandberg-Hanssen}, E. (1995).
\newblock {\em {The nature of solar prominences}}, volume 199 of {\em
  Astrophysics and Space Science Library}.

\bibitem[{Title} \& {Rosenberg}(1981)]{1981siwn.conf..326T}
{Title}, A. \& {Rosenberg}, W. (1981), Opt. Eng., 20, 815

\bibitem[{van Noort} et al.(2005)]{2005SoPh..228..191V}
{van Noort}, M., {Rouppe van der Voort}, L., \& {L{\"o}fdahl}, M.~G. (2005).
\newblock {\solphys}, {228}, 191

\bibitem[{van Noort} \& {Rouppe van der Voort}(2006)]{2006ApJ...648L..67V}
{van Noort}, M.~J. \& {Rouppe van der Voort}, L.~H.~M. (2006).
\newblock {\apjl}, {648}, L67

\bibitem[{Vial}(1998)]{1998ASPC..150..175V}
{Vial}, J.-C. (1998).
\newblock In {D.~F.~Webb, B.~Schmieder, \& D.~M.~Rust}, editor, {\em IAU
  Colloq. 167: New Perspectives on Solar Prominences}, volume 150 of {\em
  Astronomical Society of the Pacific Conference Series}, pages 175

\bibitem[{Zaqarashvili} et al.(2010)]{2010MNRAS.404L..74Z}
{Zaqarashvili}, T.~V., {Kukhianidze}, V., \& {Khodachenko}, M.~L. (2010).
\newblock {\mnras}, {404}, L74


\end{thebibliography}
\end{document}